\documentclass[journal=jacsat,manuscript=article, layout=singlecolumn]{achemso}

\usepackage[version=3]{mhchem}
\usepackage{amsmath} 
\usepackage{amssymb} 
\usepackage{amsfonts}

\usepackage{graphicx}
\usepackage{color}

\usepackage{array}
\usepackage{multirow}

\providecommand{\dd}{\mathrm{d}}                			
\renewcommand{\vec}[1]{\mbox{\boldmath$#1$}}    	

\author{Vladimir Palivec}
\affiliation{Department of Physical Chemistry, University of Chemistry and Technology, Prague, Technicka 5, 16628, Prague, Czech Republic}
\author{Denis Zadrazil}
\affiliation{Department of Physical Chemistry, University of Chemistry and Technology, Prague, Technicka 5, 16628, Prague, Czech Republic}
\author{Jan Heyda}
\affiliation{Department of Physical Chemistry, University of Chemistry and Technology, Prague, Technicka 5, 16628, Prague, Czech Republic}
\email{jan.heyda@vscht.cz}

\title{All-atom REMD simulation of poly-N-isopropyl-acrylamide thermodynamics in water: a model with a distinct 2-state behavior}

\begin{document}


\begin{abstract}
Poly(N-isopropylacrylamide) is a thermoresponsive polymer, an essential building block of a large family of soft smart materials. In neat water it undergoes upon heating  at $\approx$305\,K a phase transition from swollen to collapsed state. This interesting polymer behavior has been subject to a large number of computer modelling studies, mostly described with original OPLS force field suggested in the first works, where qualitatively correct behavior was observed. Nevertheless, employing converged replica exchange molecular dynamics simulations, we have shown on PNIPAM 30mer that this widely used force-field can not describe the phase transition thermodynamics. We found that the reason is a poor balance of hydrophobic-polar character of the polymer. We have developed and tested three new force fields OPLS1.2x, QM1, and QM2, which all exhibit two state behavior, and reasonably reproduce experimental lower critical solution temperature and transition thermodynamics. The best performance was found for QM2, which is suggested for future PNIPAM studies. Finally, the effect of polymer chain length on collapse transition thermodynamics was investigated by simulations of 20, 30 and 40mer chains. Consistent with experimental data, the transition enthalpy per monomer unit is only weakly increasing with the chain length. However, the calculated transition enthalpy is around times smaller than the experimental calorimetry value.
\end{abstract}

\section{Introduction}

Stimuli responsive soft materials already proved their potential in number of applications, including artifical tissues, biocompatible coatings, smart sensors, or work performing actuators.\cite{Stuart_NatMat,Yuan_AdvMat,Yuan_NatComm,Bajpai2008,Alarcon_ChemSocRev_2005} Vast majority of current applications is based on in depth experimental investigation of ad-hoc syntesized compounds, which were found perspective.
Among the soft materials, the prominent role plays a thermoresponsive Poly-N-isopropyl acrylamide (PNIPAM), with most detail understanding of physico-chemical behavior available.\cite{Winnik_50year_PNIPAM,Heskins1968,Philipp_SoftMatter_2013,Philipp_SoftMatter_2012,Wu_PNIPAM_DLS_PRL_2001,Wu_PNIPAM_DLS_Macromol_2007} The homopolymer undergoes rapid collapse transition at 32$^\circ$C in pure water and under most widely applied conditions.\cite{Winnik_50year_PNIPAM,Heskins1968} Despite, 50 years of research even this system possesses many challenges for the community, which may look technical at first, but are rather general and will most likely appear also in other thermoresponsive polymer families.\cite{Lutz_PNIPAM_time_over_JACS_2006,Aseyev2011} The most important stems from the role of terminal group functionalization, polydisperisty of the polymer solution, the effect of tacticity and potential branching, which are related to the polymerization conditions.

These issues ask for well defined systematic studies, which may be readily answered with the aid of atomistic molecular dynamics simulations (MD). Since the first work of PNIPAM chain in pure water, the effect of salts, osmolytes, but also tacticity was investigated.\cite{Tavagnacco_PNIPAM_PCCP_2018,Rodriguez-Ropero2015,Rodriguez-Ropero2014,Rodriguez-Ropero2015b,PNIPAM_Vrabec_FluidPhase_2010,VanderVegt2006,Tucker2012,Mukherji_PNIPAM100_SoftMatter_2016,Algaer2011} The dominant protocol consisted of number of straightforward MD simulations under designed set of conditions. The observed differences in observed transition temperature (related to LCST) were quantitatively analysed and compared to the available experimental data.\cite{Philipp_SoftMatter_2013,pnipam:lcst,cremer:2007} From the early beginning, the issues with the poor sampling of transition events become evident.\cite{PNIPAM_Vrabec_FluidPhase_2010,Algaer2011}. Fortunatelly, the separate investigation of properties of coil and globular states is still available, however, not the relative populations of collapsed and swollen states.
To bypass this issues it was suggested to perform thousands of short (on the order of tens of ns) independent simulations initiated from different polymer conformations and calculate the ensemble averages.\cite{Dalgicdir_PNIPAM_manyMDs_JPCB_2017}

Recently, the role of initial conformation on the coil-to-globule transition of 30mer PNIPAM chain was systematically studied in $\mu$s long simulations. It was found that tightly collapsed states do not reswell on the simulation timescale. This observation sets all quantitative conclusions based on MD simulations in question, since the free energy balance between the two states is hardly obtained in direct MD simulations.\cite{Kang_PNIPAM_long_MD_2016} With the increasing chain length the situation do not improve either, since the chain correlation time scales with the chain length.\cite{Mukherji_PNIPAM100_SoftMatter_2016}

The most direct thermodynamic parameter is the enthalpy of collapse of long PNIPAM chains at finite polymer concentrations as determined in differential scanning calorimetry experiments.\cite{Kato_PNIPAM_DSC_2005,Freitag_PNIPAM_DSC_Langmuir_2002,Ptitsyn1994,Ptitsyn1995,Kujawa2001,Akashi2002} These are mostly operated on approx. 500mer samples ($\approx 50$\,kDa), which are typical a product of radical polymerization. From experiments with systematically varied polymer sizes, it was proposed that the collapsing domains are roughly 100 monomer long. In order to compare the experimental data of different polymer samples, the collapse enthalpy is typically calculated per gram of PNIPAM. The collapse enthalpy grows with the increasing polymer length and saturates at around $50$\,J$\cdot$\,g.\cite{Kato_PNIPAM_DSC_2005,Ptitsyn1994,Akashi2002}

The thermodynamics comparison of simulation and experimental in soft matter is best extablished in protein denaturation. Here, good qualitative agreement is usually obtained (i.e., in term of denaturation free energy), however, the splitting into enthalpy and entropy contributions are unsatisfactory, with large absolute errors.\cite{Garcia_TrpCage_DSC_comparison_Protein_2010}

In this work, we have performed large scale replica-exchange MD simulations to calculate the equilibrium thermodynamic properties of PNIPAM in pure water for the original OPLS force field, which is abundant in literature. Next, we have proposed and tested improved PNIPAM models, which provide better agreement with the experimental data.

\section{Methods}
\subsection{General computational setup}

All molecular dynamics simulations were performed using Gromacs 5.0.4. simulation package.\cite{Abraham2015} In the first set of simulation, the investigated system contained a single 30 monomers long PNIPAM chain, described by the OPLS force field\cite{PNIPAM_Vrabec_FluidPhase_2010,Rodriguez-Ropero2014,Algaer2011,Mukherji_PNIPAM100_SoftMatter_2016} (or modified OPLS1.2x, QM1, and QM2, see below) and 6370 SPC/E water molecules\cite{SPCE}, which after equilibration resulted into a cubic box of approximatelly 6$\times$6$\times$6 nm$^3$. In the second set of simulations, we have investigated the effect of PNIPAM chain length, i.e., 20mer and 40mer, on the collapse thermodynamics. In this set of simulations, QM2 force-field was employed, and the system size with 20 and 30mer chains was kept the same, while the equilibrium box size increased to approximatelly 7$\times$7$\times$7 nm$^3$ for 40mer.

Replica exchange molecular dynamics (REMD) simulations were performed in isobaric-isothermal ensembles using 76 replicas distributed over temperatures 250-420\,K so that the exchanges were attempted every 100\,fs and the acceptance probability between the neighboring replicas was 0.25-0.3.\cite{REMD_technique_PCL2001} The temperature coupling was achieved using the velocity rescale thermostat for canonical sampling\cite{Vrescale_thermostat_Bussi} with a coupling frequency 0.1\,ps$^{-1}$, while the pressure coupling was maintained by the Parrinello-Rahman barostat\cite{Pcoupling_Parrinello_Rahman_JApplPhys_1981}, held at p = 1\,bar, with a coupling constant of 2 ps$^{-1}$. Isotropic 3D periodic boundary conditions were applied. A general cutoff 1.0\,nm was used both for Coulomb and van der Waals interactions. Neighbor list was generated using the Verlet cutoff scheme and the neighbor list was updated every 40 steps. The long range electrostatics were accounted for by the particle mesh Ewald method\cite{Berkowitz_PME_1995}. All bonds containing hydrogen atoms were constrained using LINCS algorithm\cite{LINCS_Hess1997,LINCS_Hess2008a}. Production REMD simulations were run for 100-200\,ns per replica with a time step of 2\,fs. Prior to a production run the minimization was performed using 1000 steps of steepest descend method followed by 25\,ns of equilibration.

\subsection{PNIPAM force field refinement}

To describe the PNIPAM polymer we used the OPLS forced field as referenced in the literature\cite{PNIPAM_Vrabec_FluidPhase_2010,Rodriguez-Ropero2014,Algaer2011,Mukherji_PNIPAM100_SoftMatter_2016} and its modifications, where only atomic partial charges were slightly modified (see below). Other properties (i.e., bonded and LJ parameters) of the original OPLS force field remained untouched. 

\begin{figure}[h!]
\begin{center}
\includegraphics[width=4.0cm,angle=0]{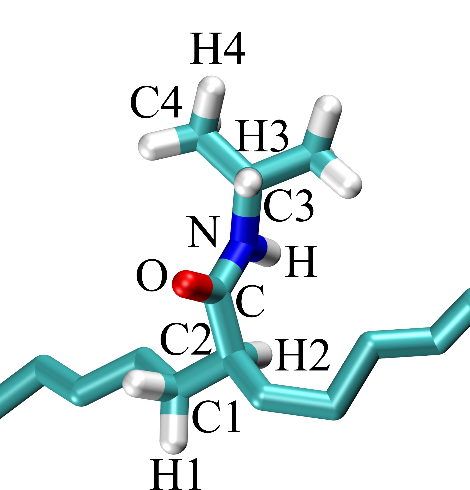}
\caption{Explanation of atom type names in the monomer unit of PNIPAM. The partial charges on individual atoms in the original OPLS force field, as well as in the modified force fields are presented in Table \ref{Table_partial_charges}.}
\label{PNIPAM_labels}
\end{center}
\end{figure}

The OPLS1.2x force field was obtained by scaling all OPLS partial atomic charges by a factor of 1.2. The QM1 force field was based on partial charge calculations from quantum mechanical calculations in Gaussian software\cite{g16}, employing HF/6-31G* quantum chemical method, and applying the RESP procedure\cite{Kollman_RESP_gaff_2004} on monomeric unit and oligomers (3mer, 5mer, and 20mer). The simulations with QM1 force field resulted in a 2-state behavior, however, the of swollen states were populated at temperatures by about 30\,K lower than experimental lower critical solution temperature (LCST). In order to increase the LCST towards the experimental value, we have introduced QM2 force field, which possesses more polar amide bond (closer to the OPLS partial atomic charges), and rest of the polymer unit is described by the QM1 atomic partial charges. Figure~\ref{PNIPAM_labels} provides the labeling of individual atoms in the monomer unit and Table~\ref{Table_partial_charges} summarizes the sets of PNIPAM partial charges (i.e., force-fields), employed in this work.

\begin{table}[htbp]
\caption{Four different force fields, namely: OPLS, OPLS1.2x, QM1, and QM2, which were used in this study to model a PNIPAM monomer unit. The atom labeling is shown in Figure~\ref{PNIPAM_labels}.}
\begin{scriptsize}
\begin{tabular}{|c|c|c|c|c|c|c|c|c|c|c|c|c|}
\hline
 & C1 & H1 & C2 & H2 & C & O & N & H & C3 & H3 & C4 & H4 \\ \hline
OPLS & -0.12 & 0.06 & -0.06 & 0.06 & 0.5 & -0.5 & -0.5 & 0.3 & 0.14 & 0.06 & -0.18 & 0.06 \\ \hline
OPLS1.2x & -0.144 & 0.072 & -0.072 & 0.072 & 0.6 & -0.6 & -0.6 & 0.36 & 0.168 & 0.072 & -0.216 & 0.072 \\ \hline
QM1 & -0.18 & 0.09 & 0 & 0.05 & 0.3 & -0.5 & -0.38 & 0.27 & 0.36 & 0.06 & -0.32 & 0.08 \\ \hline
QM2 & -0.18 & 0.09 & 0 & 0.05 & 0.5 & -0.57 & -0.57 & 0.33 & 0.36 & 0.06 & -0.32 & 0.08 \\ \hline
\end{tabular}
\end{scriptsize}
\label{Table_partial_charges}
\end{table}

\subsection{Evaluation of the collapse transition thermodynamics}
In order to evaluate the transition thermodynamics, i.e., the population of collapsed and swollen state, an indicator must be defined. In this work, we have employed a radius of gyration $R_g$, which is defined via equation \ref{Rg_definition}.

\begin{align}
\left\langle R_{g}^{2}\right\rangle &= \frac{1}{N} \sum_{i=1}^{N}(\vec{r_i}-\vec{r}_{mean})^2 \label{Rg_definition}
\end{align}

We stress that a REMD simulation provides us with equilibrated ensemble over a broad range of temperatures, thus we have access to a 2-dimensional probability distribution $P(R_g, T)$. This clearly visualizes the underlaying thermodynamic landscape.

To distinguish between the two polymer states, we have to define a threshold value, $R_{g,0}$. As a suitable $R_{g,0}$ value we have chosen a point, where the polymer size distributions at low (below LCST) and high (above LCST) temperature cross, see SI for more details. These values are summarized in Table \ref{Rg0_summary}. Note that other rational choices of polymer state descriptors would work equally well. The population of polymer in a collapsed ($P_c$) and swollen state ($P_s$) are defined by equation~\ref{Pcs}.

\begin{table}[htbp]
\caption{The threshold values, $R_{g,0}$, defined for all investigated force-fields and chain lengths, which are applied to define the population of collapsed and swollen states of the polymer.}
\begin{tabular}{|c|c|r|}
\hline
PNIPAM force field & system & \multicolumn{1}{c|}{$R_{g,0}$ [nm]} \\ \hline
OPLS & 30mer & N.D. \\ \hline
OPLS1.2x & 30mer & 1.3 \\ \hline
QM1 & 30mer & 1.45 \\ \hline
QM2 & 30mer & 1.33 \\ \hline
QM2 & 20mer & 1.15 \\ \hline
QM2 & 40mer & 1.55 \\ \hline
\end{tabular}
\label{Rg0_summary}
\end{table}

\begin{align}
P_{c}(T) &= \int_{0}^{R_{g,0}}  P(R_g; T) \dd R_g \label{Pcs}  \\
P_{s}(T) &= \int_{R_{g,0}}^{\infty} P(R_g; T) \dd R_g \nonumber
\end{align}

The equilibrium constant, $K(T)$, and transition free energy of unfolding, $\Delta G_u(T)$ are defined by equations \ref{K}, and \ref{DG}, where $R$ is the universal gas constant, and $T$ is the absolute temperature. The underlaying transition enthalpy, $\Delta H_{0}$, entropy, $\Delta S_{0}$, constant pressure heat capacity, $\Delta C_{p,0}$, and the transition temperature $T_0$ were obtained via fitting of equation~\ref{DGfit} toMD data. Note that the transition temperature $T_0$ is obtained from the condition that $\Delta G_u(T_0) = 0$, and thus $\Delta H_0 = T_0 \Delta S_0 $.

\begin{align}
K(T)        &= \frac{P_{s}(T)}{P_{c}(T)} \label{K}  \\
\Delta G_u(T) &= -RT \ln K(T)              \label{DG}  \\
\Delta G_u(T) &= \Delta H_0 - T \Delta S_0 + \Delta C_{p,0} \left( T-T_0 - T \ln \frac{T}{T_{0}} \right) \label{DGfit}
\end{align}

\clearpage
\newpage
\section{Results}

\subsection{Force-field performance on thermodynamics of PNIPAM 30mer}

\begin{figure}[h!]
\begin{center}
\includegraphics[width=12.0cm,angle=0]{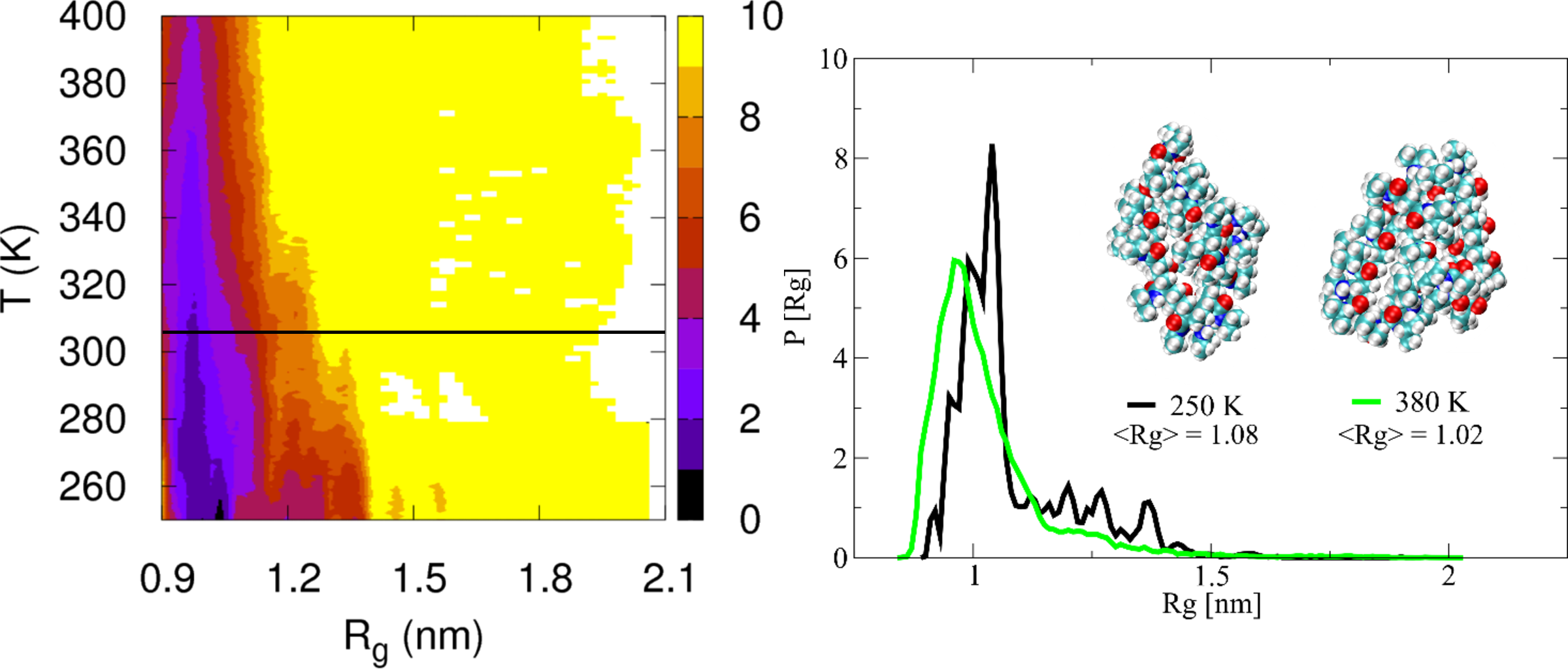}
\caption{Distribution of polymer size (radius of gyration) for PNIPAM 30mer as a function of temperature, described with the original OPLS force field and sampled in a REMD simulation. Left, 2-dimensional free energy landscape (kJ/mol), $PMF(R_{g},T)=-RT \ln P(R_{g},T)$, of the polymer size over the whole investigated temperature range, where the black horizontal line highlights the experimental LCST. Note that other than fully collapsed states start to be marginally populated only at temperatures much lower than 270\,K. Right, polymer size distribution at selected temperatures, i.e., below (250\,K) and above (380\,K) the experimental critical temperature. Typical polymer conformations with the highest probability are shown in the inset.}
\label{OPLS_poor_sampling}
\end{center}
\end{figure}

First, we have performed REMD simulations of PNIPAM 30mer, employing the standard OPLS force field, which application dominates in recent literature. Our aim, similar to that in protein denaturation studies, was to quantify the population of collapsed and swollen polymer states and consequently to evaluate the transition thermodynamics. In particular, we are interested in the experimental properties, such as transition enthalpy $\Delta H(T_{0})$ and temperature $T_{0}$, which are available from calorimetric and cloud point studies. Figure \ref{OPLS_poor_sampling} (left) presents the free energy landscape in terms of the polymer size ($R_g$) and the temperature in the range of 250-400\,K. Figure \ref{OPLS_poor_sampling} (right) shows polymer size distributions below (250\,K) and above (380\,K)the critical temperature $T_0$, including the representative structures.

\begin{figure}[h!]
\begin{center}
\includegraphics[width=10.0cm,angle=0]{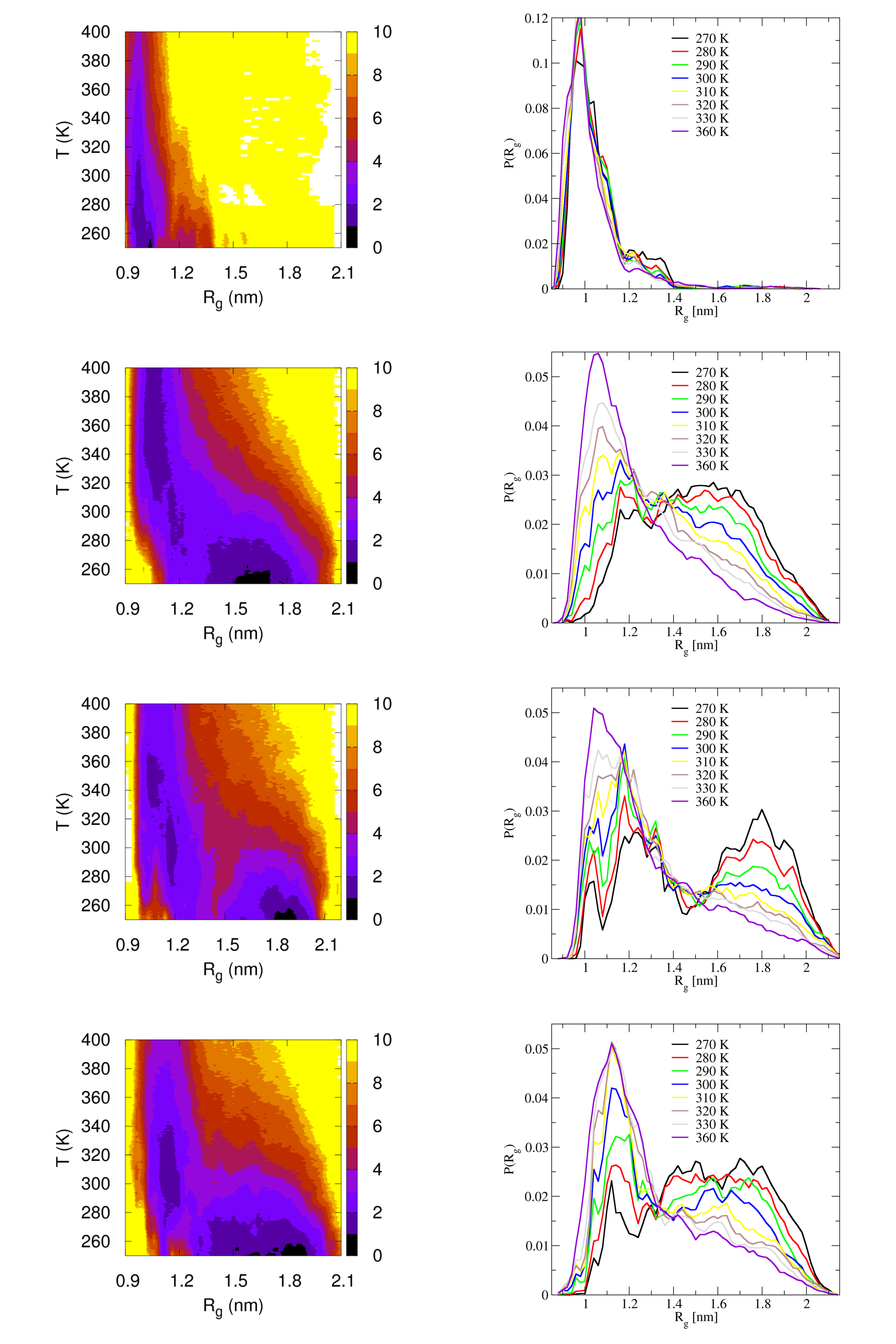}
\caption{Distribution of the polymer size (radius of gyration) for PNIPAM 30mer, modeled with OPLS (top row), OPLS1.2x (second row), QM1 (third row), and QM2 (bottom row) force fields and sampled in REMD simulations. 2-dimensional free energy landscape (kJ/mol) is presented on the left, i.e., the potential of mean force $PMF(R_{g},T)=-RT \ln P(R_{g},T)$ of the polymer size over the whole investigated temperature range. The probability distribution of polymer sizes at selected temperatures (in 10\,K steps) around the LCST are presented on the right.}
\label{OPLS_vs_new_models}
\end{center}
\end{figure}

Notably, no partially ($R_{g} \approx 1.5{\rm\,nm}$) or even completelly swollen PNIPAM states were populated with the OPLS force-field. Although the distribution weakly broadens with decreasing temperature (i.e., around 250\,K), the simulation fails to reproduce the experimental reality. We have thus hypothesized that within the OPLS force-field, the balance between polar and hydrophobic character of the monomer unit is too inaccurate. In order to increase the polarity, we have uniformly scaled the partial charges in the monomer unit. Employing other REMD simulations (see SI), we have found that OPLS1.2x is a reasonable modification of the original force-field. Independently, we have tested ab-initio derived partial charges QM1 and the final version QM2. The results are summarized in Figure \ref{OPLS_vs_new_models}, showing that a two-state behavior was achieved by all refined models.

\begin{figure}[h!]
\begin{center}
\includegraphics[width=10.0cm,angle=0]{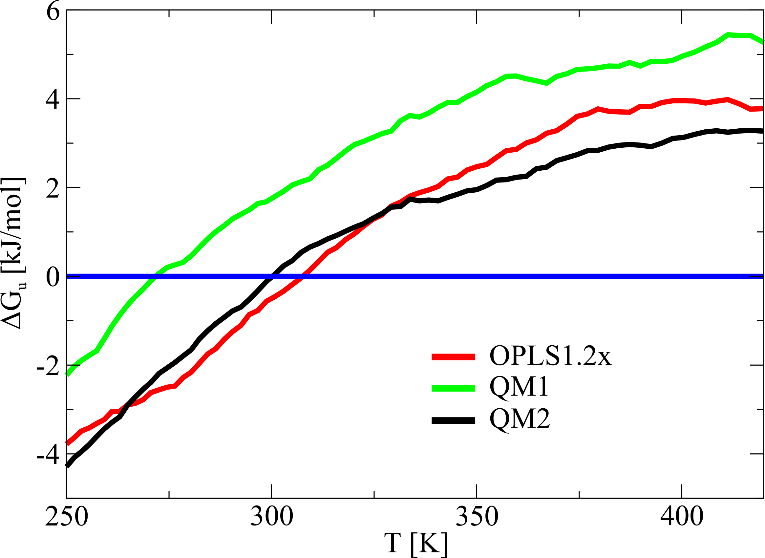}
\caption{Summary of transition thermodynamics for PNIPAM force field, which were developed in this work. The transition free energy is determined from the populations of collapsed and swollen states, i.e., from the polymer size distribution presented in Figure \ref{OPLS_vs_new_models}. Note that for the original OPLS, no swollen-to-collapse transition could be determined. Thermodynamic parameters were obtained by fitting this rough data, and are summarized in Table \ref{Thermodynamics_tested_models_table}.}
\label{DGu_tested_models}
\end{center}
\end{figure}

The data were further analyzed as stated in Methods section and thermodynamic descriptors, such transition free enegry, temperature, enthalpy, entropy and heat capacity for PNIPAM 30mer, were evaluated. The results are presented in Figure \ref{DGu_tested_models} and Table \ref{Thermodynamics_tested_models_table}. Clearly, in all cases, the population of swollen state decreases with increasing temperature, and the transition temperature (i.e., temperature, where $\Delta G_u(T_0) = 0$) is near 300\,K.

\begin{table}[htbp]
\caption{Thermodynamic parameters of PNIPAM 30mer result from fitting of the transition free energy of swelling, $\Delta G_{u}(T)$, in Figure \ref{DGu_tested_models}. The raw data were obtained from REMD simulations  with different force-fields that are employed in this work. The energies are expressed per PNIPAM-30mer molecule.}
\begin{tabular}{|c|c|c|c|c|c|}
\hline
parameterization & $T_0$ & $\Delta H_0$ & $\Delta S_0$ & $\Delta C_{p,0}$ \\ \hline
 & [K] & [kJ\,mol$^{-1}$] & [J\,mol$^{-1}$\,K$^{-1}$] & [J\,mol$^{-1}$\,K$^{-1}$] \\ \hline
OPLS  & N.D. & N.D. & N.D. & N.D. \\ \hline
OPLS1.2x & 305 & $-20$ & $-67.2$ & 156.7 \\ \hline
QM1 & 274 & $-17$ & $-62.2$ & 186.7 \\ \hline
QM2 & 303 & $-20$ & $-65.0$ & 204.9 \\ \hline
exp.& 307 & $-110$ to $-150$ & $-360$ to $-480$ & N.D. \\ \hline
\end{tabular}
\label{Thermodynamics_tested_models_table}
\end{table}

All force-fields lead to qualitatively similar thermodynamics, however, the critical temperature $T_0$ of OPLS1.2x and QM2 is closest to the experimental value. Based on more rational values of atomic partial charges, we have selected QM2 force-field for further use. In the next section, the effect of chain length on transition thermodynamics of PNIPAM is investigated.

\subsection{Effect of chain length on the thermodynamics of PNIPAM}

In order to quantitatively determine the effect of PNIPAM chain length on the transition thermodynamics, we have employed REMD simulations of 20mer, 30mer, and 40mer PNIPAM chains (QM2 force-field) in exmplicit water.

\begin{figure}[h!]
\begin{center}
\includegraphics[width=10.0cm,angle=0]{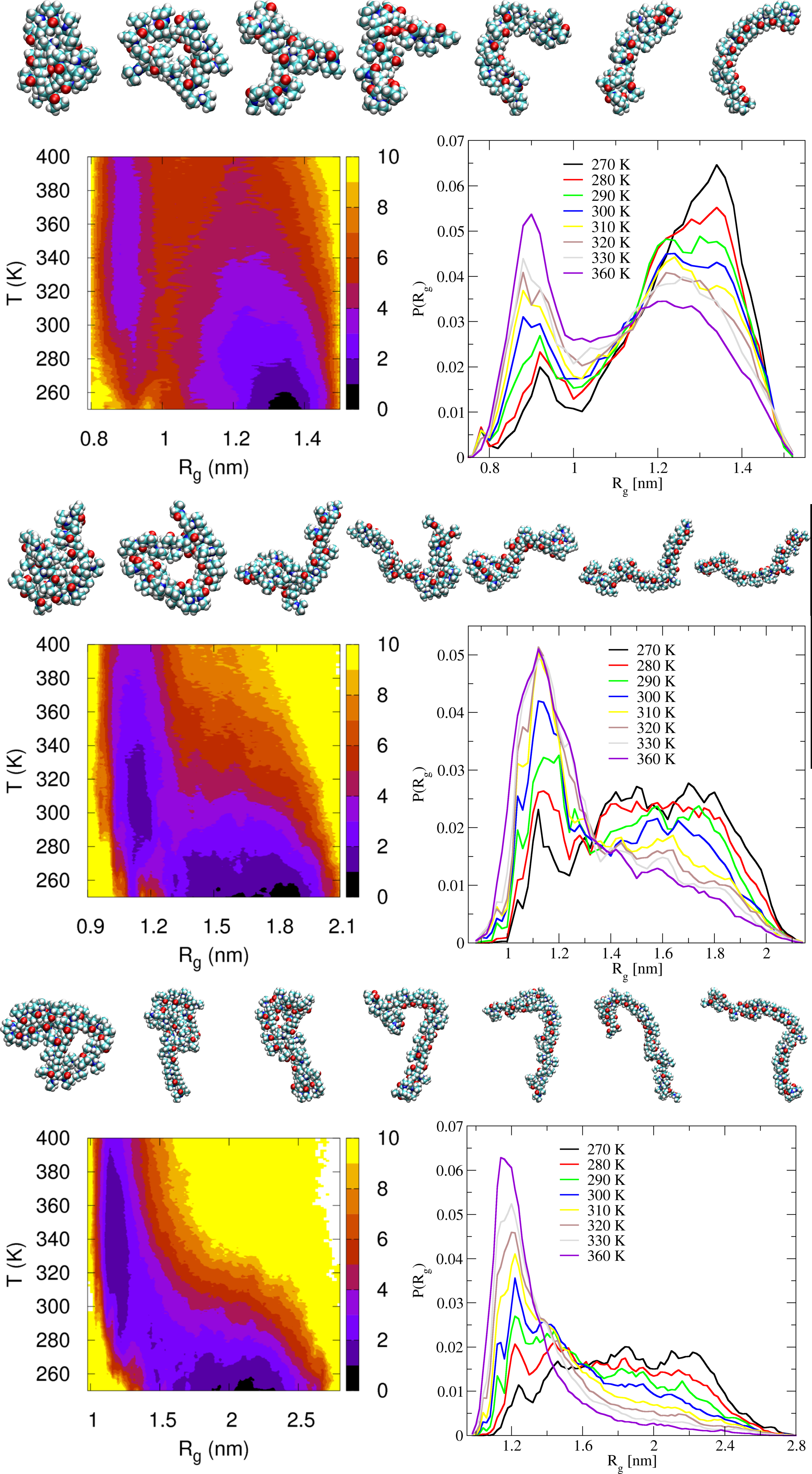}
\caption{Summary of transition thermodynamics for PNIPAM polymer of different chain lengths (20, 30, 40mer), obtained from REMD simulations, employing QM2 force-field. Left panel presents 2-dimensional potential of mean force (kJ/mol), $PMF(R_{g},T)$, while the right panel presents 1-dimensional probability distribution of polymer sizes at selected temperatures (in 10\,K~steps) around the LCST. For each polymer chain length, we present seven polymer conformations which are equally spaced in the explored R$_g$ range.}
\label{PMF_chain_legth}
\end{center}
\end{figure}

Figure \ref{PMF_chain_legth} presents the population of collapsed to swollen polymer states over the temperature range (250\,K-400\,K). In the left panels of Figure \ref{PMF_chain_legth} we present the 2D-free energy landscape, which documents the global depopulation of swollen states with the increasing temperature. The polymer size distributions at selected temperatures are presented in the right panels. It is noteworthy that while the 20 and 30\,mer exhibit a barrier between the two states (which may correspond to the transition state), the barrier is almost absent for 40\,mer. Thus the barrier is diminishing with increaing chain length and the polymer size distribution significantly broaden around the LCST, which is another indication of the critical region vicinity.
Moreover, we have found that for 30 and 40mer the swollen state could be almost completely depopulated at sufficiently high temperature. The situation is different for 20mer chain, where a significant fraction of the polymer is not collapsed, even when reaching temperatures well above the LCST. This may be related to almost absent interior of the collapsed state for too short PNIPAM chains, which may play an important role as a stabilization seed of collapsed conformations.

\begin{figure}[h!]
\begin{center}
\includegraphics[width=10.0cm,angle=0]{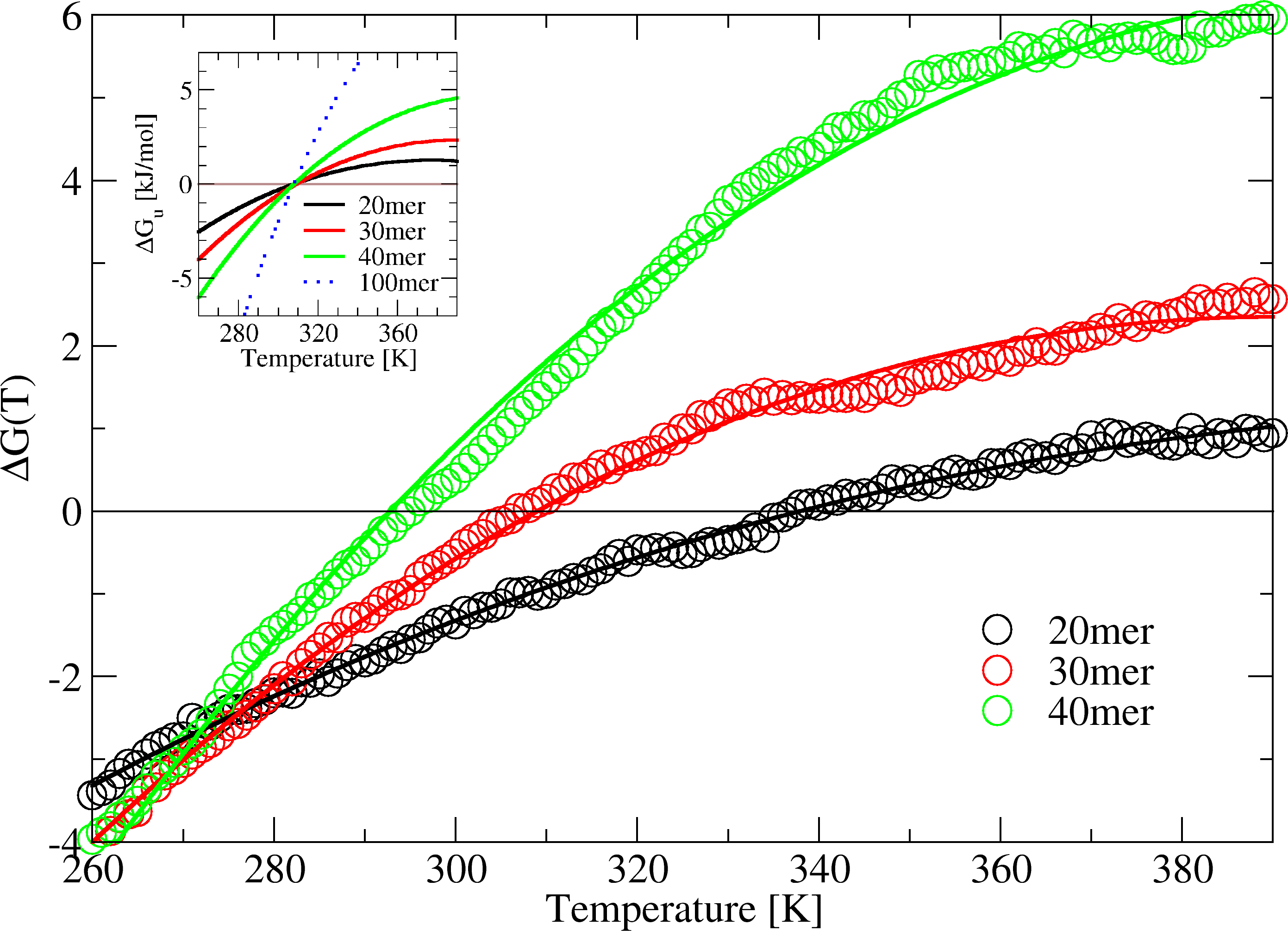}
\caption{Summary of transition thermodynamics of PNIPAM for chains, which consist of 20, 30, and 40mer. The transition free energy is determined from the populations of collapsed nad swollen state, i.e., from polymer size distribution, presented in Figure \ref{PMF_chain_legth}, and calculated from Equation \ref{Pcs}. Thermodynamic parameters obtained by fitting Equation \ref{DGfit} to the data, are summarized in Table \ref{Thermodynamics_chain_length}. The inset presents the sharpness of the collapse transition, when all the curves are offset to the same transition temperature. The experimentally suggested single domain, i.e., $\Delta G_{u}(T)$ of 100\,mer is also included (see the dotted line).}
\label{DGu_chain_length_effect}
\end{center}
\end{figure}

Analysis was performed as previously described and the transition free energy as a function of temperature for 20, 30, and 40mer chains are summarized in Figure \ref{DGu_chain_length_effect}. First, the slope of $\Delta G_{u}(T)$, is increasingly steeper for longer chains, which becomes evident when $\Delta G_{u}$ curves are offset to the same temperature (see the inset). Moreover, upon the fitting of Equation \ref{DGfit}, we have found that even the transition enthalpy per monomer increases with the chain length, see Table~\ref{Thermodynamics_chain_length}. Consequently the enthalpy of 40\,mer is almost 4x that of 20\,mer, pointing to significant contribution of many-body effective interactions, when short polymer chains at low concentrations are assumed.

\begin{table}[htbp]
\caption{Chain length effect on thermodynamic parameters of PNIPAM. Results for 20, 30, and 40mer PNIPAM chains are obtained by the fitting of the Equation~\ref{DGfit} to transition free energy of swelling, $\Delta G_{u}(T)$, presented in Figure \ref{DGu_chain_length_effect}. Thermodynamics parameters per polymer chain and per monomer are shown to see the chain-length effect better. All values were obtained from REMD simulations with QM2 force-field.}
\begin{tabular}{|c|c|c|c|c|}
\hline
chain & $T_0$ & $\Delta H_0$ & $\Delta S_0$ & $\Delta C_{p,0}$  \\ \hline
 & [K] & [kJ\,mol$^{-1}$] & [J\,mol$^{-1}$\,K$^{-1}$] & [J\,mol$^{-1}$\,K$^{-1}$]\\ \hline
20mer & 338 & -9.5    & -28  & 120 \\ \hline
30mer & 300 & -18.6   & -60  & 260 \\ \hline
40mer & 293 & -33.7   & -115 & 340 \\ \hline
\hline
chain & $T_0$ & $\Delta H_0$/mon.  & $\Delta S_0$/mon. & $\Delta C_{p,0}$/mon. \\ \hline
 & [K] & [kJ\,mol$^{-1}$] & [J\,mol$^{-1}$\,K$^{-1}$] & [J\,mol$^{-1}$\,K$^{-1}$]\\ \hline
20mer & 338 & -0.48 & -1.4 & 6.0 \\ \hline
30mer & 300 & -0.62 & -2.0 & 8.7 \\ \hline
40mer & 293 & -0.84 & -2.9 & 8.5 \\ \hline

\end{tabular}
\label{Thermodynamics_chain_length}
\end{table}

\section{Conclusion}

In light of number of PNIPAM molecular dynamics simulations, reported in the literature,\cite{PNIPAM_Vrabec_FluidPhase_2010,Rodriguez-Ropero2014,Algaer2011,Mukherji_PNIPAM100_SoftMatter_2016,Stevens_Tdep_PNIPAM_JPCB_2015,PNIPAM_Vrabec_JPCB_2012} it seems rather surprising that there were no attempts to determine the PNIPAM collapse thermodynamics by well established REMD simulation method.\cite{Garcia_TrpCage_DSC_comparison_Protein_2010,REMD_technique_PCL2001} In this work, we have performed REMD simulations of a single PNIPAM 30mer chain employing the widely-used OPLS force-field. We have found that this force-field completely fails to describe the swollen states in equilibrium with the collapsed state. Consequently, the LCST cannot be determined and the description of the transition thermodynamics is meaningless. We have found that the OPLS force-field overestimates the hydrophobic character of PNIPAM with respect to its hydrophilicity, a fact that leads to the sole population of collapsed states even at temperatures as low as 250\,K. 

By adjusting atomic partial charges, we have achieved two state behavior of the polymer and were able to tune the lower critical solution temperature close to the experimental value. Of the three promising force fields, two (OPLS1.2x and QM2) correctly reproduce the lower critical solution temperature. Calculated thermodynamic properties, e.g., enthalpy of collapse, are for PNIPAM 30mer, however, approximatelly 6x times smaller than the experimental value,\cite{Kato_PNIPAM_DSC_2005,Ptitsyn1994,Akashi2002} although at finite polymer concentrations. To that end, we have evaluated the effect of polymer chain length on the collapse thermodynamics, performing the REMD simulations of a single 20, 30, and 40mer PNIPAM chain. It was found that within the investigated chain lengths the transition enthalpy increases more than linearly, which presumably originates from the absence of the polymer interior of the collapsed state for the shorter polymer chains at infinite dilution and is supported by geometries presented in Figure \ref{PMF_chain_legth}. These many-body contributions, and formation of aggregation seeds, seem to be important for a quantitative evaluation of the collapse thermodynamics of experimentally relevant polymer chain lengths.
The recommended QM2 force field was rigorously derived from quantum mechanical calculation of partial charges, and referenced to reasonable LCST and transition thermodynamics. We believe that our force-field allows to answer many urgnet questions about revesible collapse transitions of PNIPAM in aqueous solutions.




\begin{acknowledgement}
J. H. thanks the Czech Science Foundation (grant 16-24321Y) for support. This work was supported by The Ministry of Education, Youth and Sports from the Large Infrastructures for Research, Experimental Development and Innovations project 'IT4Innovations National Supercomputing Center -- LM2015070' (projects OPEN-7-50, OPEN-10-36).
\end{acknowledgement}

\clearpage
\newpage

%
%

\clearpage

\bibliography{AA_Manuscripts-PNIPAM-REMD-FF}

\end{document}